\documentclass[12pt]{iopart}
\usepackage{epsfig}  
\usepackage{cite}  

\newcommand{\pom}{\rm I\!P}

\newcommand{\alphareg}{\alpha_{_{\rm I\!R}}}

\newcommand{\xpom}{\xi}
\newcommand{\mx}{M_{_X}}

\newcommand{\gapprox}{\stackrel{>}{_{\sim}}}
\newcommand{\lapprox}{\stackrel{<}{_{\sim}}}

\newcommand{\pperp}{p_{_T}}
\newcommand{\ptj}{p_{_T}^{\rm jet}}
\newcommand{\ptrans}{p_{_T}}

\begin{document}


\title[Low $x$ and Diffraction Experimental Summary]{Recent Low 
{\boldmath $x$} and Diffractive Collider Data}

\author{P. R. Newman\footnote{Supported by the UK Particle Physics and 
Astronomy Research Council (PPARC).}}

\address{School of Physics and Astronomy, University of Birmingham, 
B15 2TT, UK. \\
E-mail: prn@hep.ph.bham.ac.uk}

\begin{abstract}

Selected recent data from collider experiments pertaining to the 
understanding of QCD at low Bjorken-x are reviewed.
The status of QCD and Regge factorisation in hard 
diffractive interactions is
discussed in terms of data from HERA and the Tevatron.
The possibility of anomalous behaviour in the $\gamma \gamma$ 
total cross section 
is confronted with the most recent measurements from LEP. Data
from all three colliders that are sensitive to possible BFKL effects 
are presented and different interpretations are discussed.

\end{abstract}




\section{Introduction}

Since the HERA and Tevatron colliders have been operational, abundant data
have become available that are sensitive to
proton structure at low parton-$x$. Data on photon structure
from HERA and LEP have been similarly impressive. This latest
generation of colliders has pushed back 
the limits of our understanding of QCD considerably. 
There is not space here to do justice to all low $x$ data. Instead,
three particularly topical areas that were discussed at the 
1999 Durham Phenomenology Workshop are singled out.

\section{Factorisation in Hard Diffraction}

The HERA and Tevatron experiments have now produced abundant high quality 
data on the `single diffractive' 
processes $\gamma^* p \rightarrow X p$ and 
$\bar{p} p \rightarrow X \! \! \! \stackrel{\hspace*{0.1cm} _{_{ (-)}}}{p}$ 
at low momentum transfer. 
Hard scales, provided for example by a highly virtual photon 
(in the $\gamma^* p$ case) or final state 
jets or heavy quarks, encourage the use of perturbative QCD as a
tool with which to understand the parton level dynamics. 
The development of techniques which simultaneously describe the
$\gamma^* p$ and $\bar{p} p$
single dissociation processes is a major
current issue in hadron phenomenology.

\begin{figure}[htb]
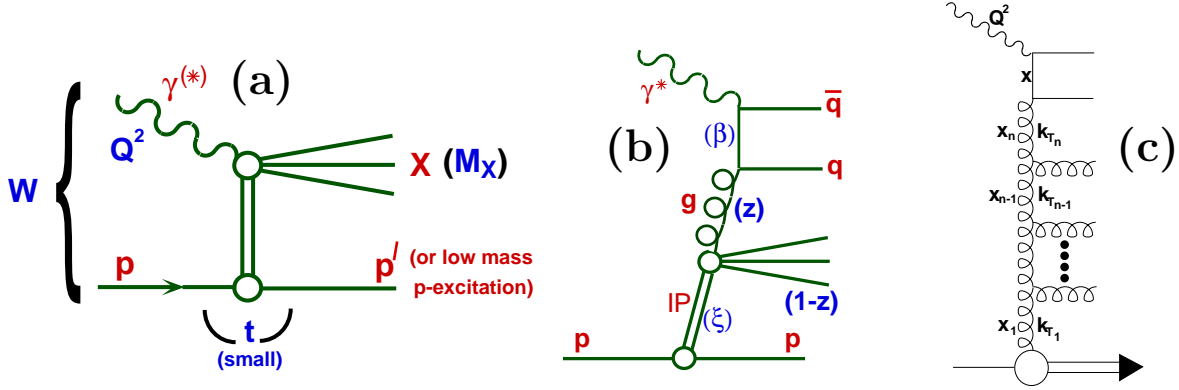
 \unitlength 1mm
    \begin{center}
    \end{center}
    \begin{picture}(120,43)
       \put(0,0){\epsfig{file=gendis.epsf,width=0.45\textwidth}}
       \put(74,0){\epsfig{file=bgf3.epsf,width=0.24\textwidth}}
       \put(125,-2){\epsfig{file=lowx2.epsf,width=0.17\textwidth}}
       \put(30,37){\Large{\bf{(a)}}}
       \put(80,28){\Large{\bf{(b)}}}
       \put(148,28){\Large{\bf{(c)}}}
    \end{picture}
    \caption{(a) The generic photon dissociation process at HERA. 
(b) Diagram of the dominating leading order
QCD process in models involving a pomeron with partonic sub-structure.
A $q \bar{q}$ pair is produced
via photon-gluon-fusion ($\gamma^* g \rightarrow q \bar{q}$).
(c) Illustration of the low $x$ parton ladder in DIS.}
    \label{diagrams}
\end{figure}

The generic diffractive process at HERA of the type $ep \rightarrow eXp$
is illustrated in figure~\ref{diagrams}a. A photon of virtuality $Q^2$
interacts with a proton at a $\gamma^* p$ invariant mass $W$ and squared
four momentum transfer $t$ to produce a dissociating photon system
$X$ of invariant masses $\mx$, the proton remaining intact. 
In the corresponding process at the Tevatron, the photon is replaced by an
anti-proton, with either of the beam particles dissociating.Two further 
variables are usually introduced; the fraction of the proton momentum that is 
exchanged to the system $X$ is denoted 
$\xi$\footnote{At HERA, $\xpom$ is usually referred to
as $x_{_{I\!\!P}}$. Here $\xi$ is used to make explicit the correspondence
with the equivalent variable at the Tevatron.}, whilst
$\beta = x / \xpom$ is the fraction of the exchanged momentum carried by the 
quark coupling to the photon. 

A QCD factorisation theorem has recently been proved for a general class
of semi-inclusive processes in 
deep-inelastic scattering (DIS), which include the single diffractive
process \cite{collins}. This implies that a 
concept of `diffractive parton distributions'
can be introduced \cite{diffpart}, expressing proton parton probability
distributions under the constraint of an intact final state
proton with particular values of $\xpom$ and $t$. The cross section for
diffractive DIS can then be expressed as
\begin{eqnarray*}
  \sigma^{\gamma^* p \rightarrow X p} (\xpom, t, x, Q^2) \sim
\sum_i f_{i/p} (\xpom, t, x, Q^2) \otimes \hat{\sigma}_{\gamma^* i} (x, Q^2) 
\ ,
\end{eqnarray*}
where $f_{i/p} (\xpom, t, x, Q^2)$ are the diffractive parton distributions,
evolving with $x$ and $Q^2$ according to the DGLAP equations at fixed $\xpom$
and $t$, and $\hat{\sigma}_{\gamma^* i} (x, Q^2)$ are parton interaction
cross sections.

The phenomenology of soft hadronic interactions suggests that it is possible
to introduce a universal factorisable pomeron exchange with a flux factor 
dependent only on $\xpom$ and $t$. 
With this additional assumption of `Regge factorisation', the framework of 
diffractive
parton distributions can be used to define parton distributions for the 
pomeron \cite{IS}, which should describe all hard diffractive scattering
processes. The diffractive DIS cross section can then be written as
\begin{eqnarray*}
  \sigma^{\gamma^* p \rightarrow X p} (\xpom, t, \beta, Q^2) \sim
f_{\pom / p} (\xpom, t) \otimes \sum_i f_{i/\pom} (\beta, Q^2) \otimes
\hat{\sigma}_{\gamma^* i} (\beta, Q^2) \ .
\end{eqnarray*}

The validity of this second hypothesis for diffractive DIS, incorporating
both QCD and Regge factorisation, has been extensively tested at HERA. 
Measurements of the total cross section for diffractive deep-inelastic 
scattering, usually presented in the form of a $t$-integrated diffractive 
structure function
$F_2^{D(3)} (\beta, Q^2, \xpom)$ \cite{H1:F2D3,ZEUS:F2D3,F2D:conf} 
have shown that, to
the present level of accuracy, the factorisation between the $\xpom$
and the ($\beta$,$Q^2$) dependence is obeyed.\footnote{The deviations
from this factorisation shown to be present in \cite{H1:F2D3}
can be explained in full when a sub-leading exchange
($f$, $\omega$, $\rho$ and / or $a$ trajectory) is introduced.} 
Parton distributions for the pomeron have been 
extracted \cite{H1:F2D3,ACTW} from
QCD analyses of the $\beta$ and $Q^2$ dependence of $F_2^D$ using the
DGLAP evolution equations. All such extractions yield parton distributions
which are heavily dominated by gluons at low scales, the gluon density 
remaining large even at high fractional momentum. Figure~\ref{diagrams}b
then represents the dominant process at leading order of QCD. A gluon
carrying a fraction $z$ of the pomeron momentum undergoes 
boson-gluon fusion ($\gamma^* g \rightarrow q \bar{q}$) with the virtual
photon. 

\begin{figure}[htb] \unitlength 1mm
    \begin{center}
    \end{center}
    \begin{picture}(120,69)
       \put(0,0){\epsfig{file=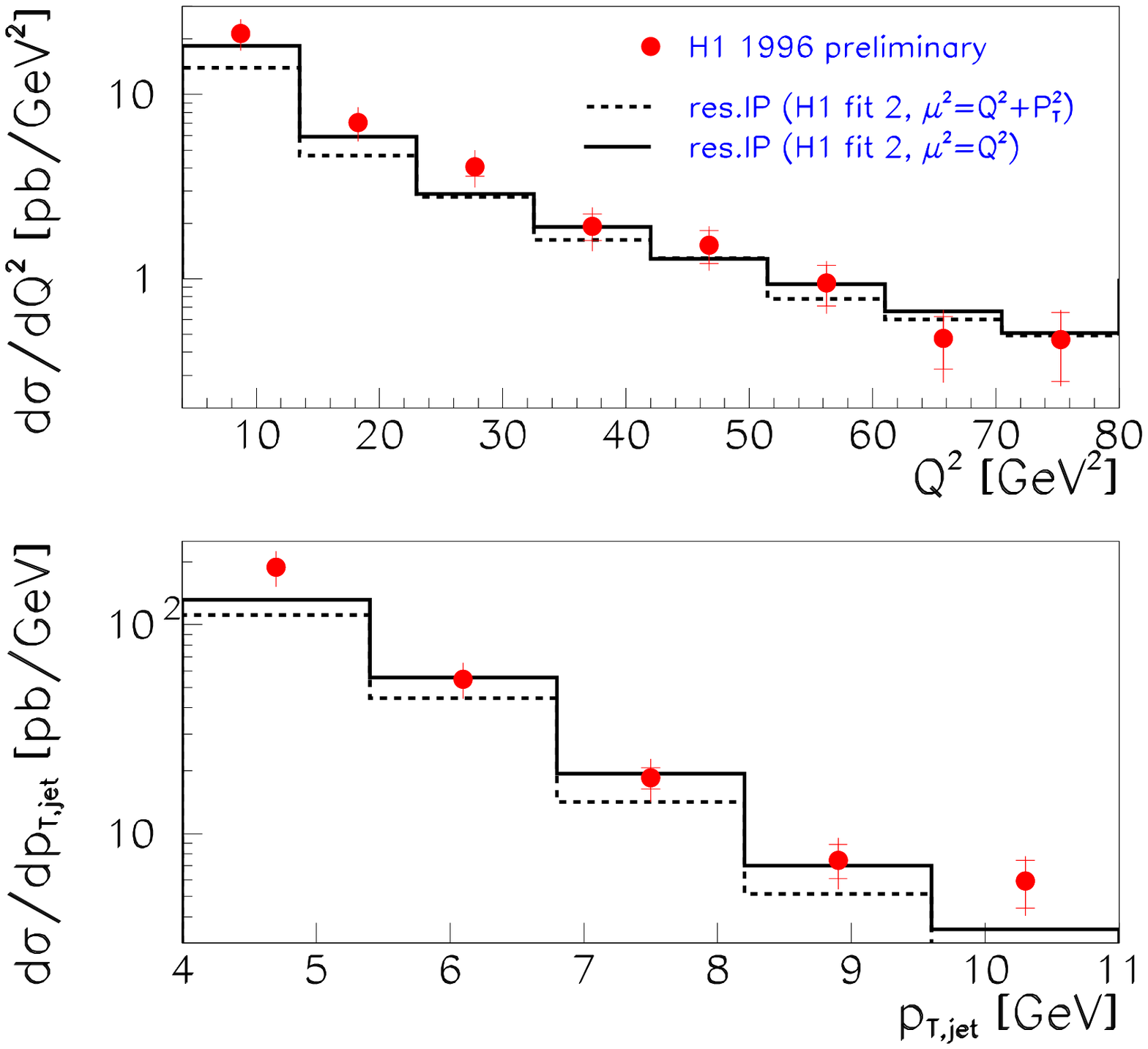,width=0.50\textwidth}}
       \put(80,-25){\epsfig{file=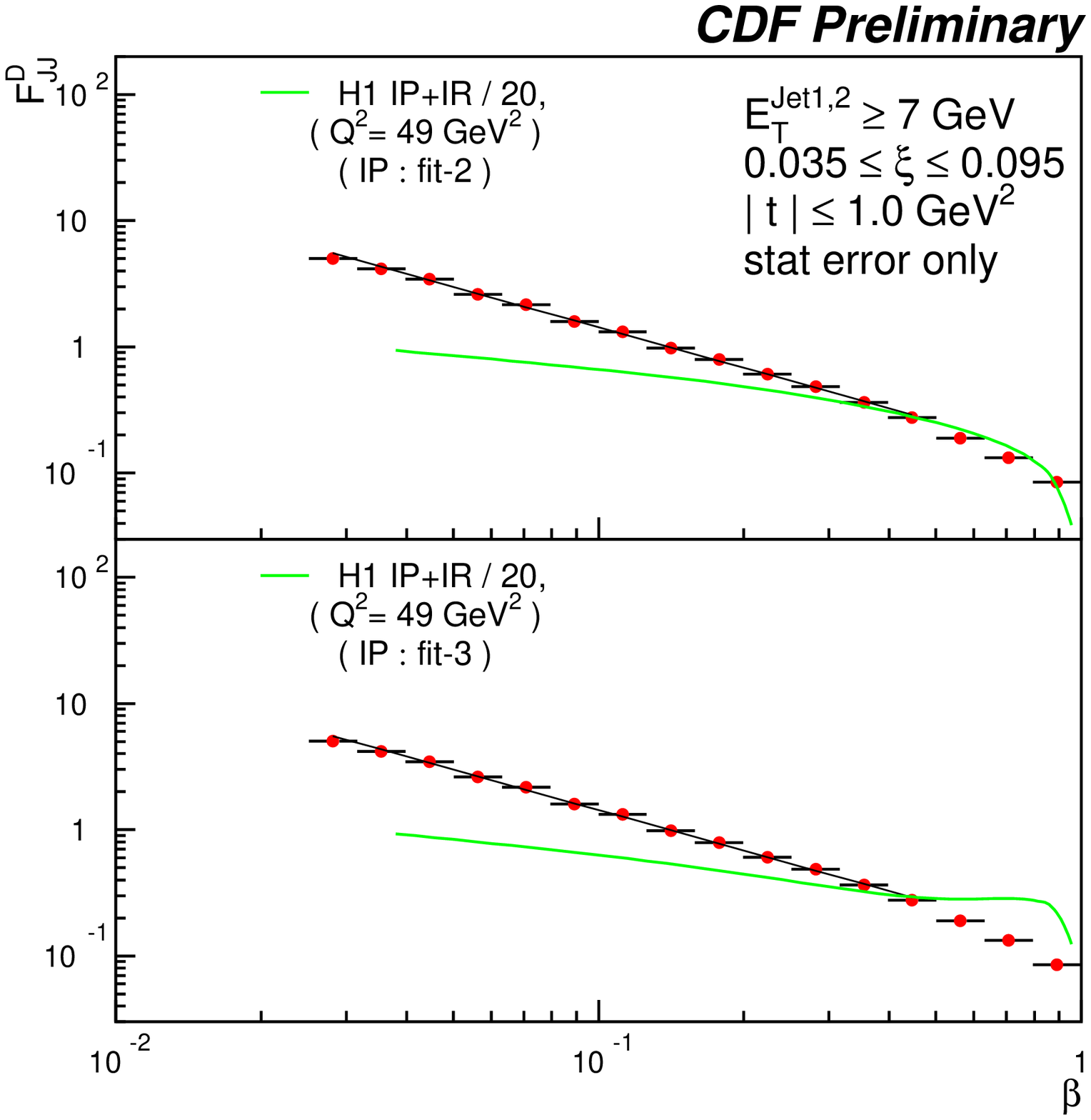,width=0.43\textwidth}}
       \put(15,15){\Large{\bf{(a)}}}
       \put(95,15){\Large{\bf{(b)}}}
    \end{picture}
    \caption{(a) $Q^2$ and jet transverse momentum ($\ptj$) distributions
for diffractive dijet electroproduction data, compared with the predictions
of the RAPGAP Monte Carlo model incorporating a set of pomeron parton 
distributions extracted from $F_2^D$ data.
(b) $\beta$ dependence of the quantity $F_{\rm JJ}^D$ (see text), related
to the fraction of $p \bar{p}$ dijet events that are produced 
diffractively. The data are compared with predictions based on two
slightly different sets of diffractive parton densities extracted from
$F_2^D$ (top and bottom plots), the predictions being reduced by a factor
of 20 for the plots.} 
    \label{jets:charm}
\end{figure}

Monte Carlo models based on the parton distributions extracted from $F_2^D$
describe HERA diffractive final state data well \cite{thrust,flow}.
The most stringent
tests come from diffractive dijet and open charm cross sections, as both are 
sensitive to the magnitude as well as the shape of the gluon distribution. 
Recent data from H1 on diffractive dijet 
electroproduction \cite{frank} are shown in figure~\ref{jets:charm}a.
The factorisable partonic pomeron model
(labelled ``res $\pom$'') gives a reasonable description
of the measurement. Similarly good agreement is
found with ZEUS data on diffractive charm 
electroproduction \cite{zeus:charm}, though a diffractive 
$D^*$ measurement from H1 in a slightly different 
kinematic region suggests deviations from
factorisation \cite{H1:charm}. With this single exception, HERA data support
the hypothesis that both QCD and Regge factorisation can be applied to all
hard diffractive processes in DIS.
 
There are good reasons to believe that the QCD factorisation theorem for 
diffractive DIS \cite{collins} cannot be extended to hard diffraction in
hadron-hadron interactions \cite{nonfac}. 
The factorisation hypothesis for $p \bar{p}$ scattering
has now been tested in some 
detail by taking parton distributions extracted from $F_2^D$ data at HERA
and using them to predict cross sections for hard diffractive processes at 
the Tevatron. By now it is clear that this approach universally predicts
cross sections well in excess of those measured. 

One example \cite{tevjets} 
is a measurement of the fraction 
$N^{\rm diff} / N^{\rm incl}$ of all dijet events that arise from the 
single dissociation process $\bar{p} p \rightarrow X p$, where the intact
final state proton has 
$0.035 < \xi < 0.095$ and is 
scattered at $|t| < 1 \ {\rm GeV^2}$. From this ratio, the quantity
\begin{eqnarray*}
    F_{\rm JJ}^D = \frac{N^{\rm diff}}{N^{\rm incl}} (x_{\bar{p}})
\left\{ x_{\bar{p}} g (x_{\bar{p}}) + \frac{4}{9} 
\left[ q (x_{\bar{p}}) + \bar{q} (x_{\bar{p}}) \right] \right\}_{\bar{p}} 
\end{eqnarray*}
is formed, where $x_{\bar{p}}$ is the Bjorken scaling variable for the 
antiproton and $x_{\bar{p}} g (x_{\bar{p}}) + \frac{4}{9} 
\left[ q (x_{\bar{p}}) + \bar{q} (x_{\bar{p}}) \right]$ represent the 
(known) effective
parton densities in the anti-proton after allowing for the leading order
colour factor of $4/9$. Assuming factorisation is valid, the resulting
quantity should correspond to the effective parton densities
of the pomeron;
\begin{eqnarray*}
    F_{\rm JJ}^D = \left\{ \beta g (\beta) + \frac{4}{9} 
\left[ q (\beta) + \bar{q} (\beta) \right] \right\}_{\pom} \otimes
f_{\pom / p} (\xpom) \ .
\end{eqnarray*}
The quantity $F_{\rm JJ}^D$ is shown as a function of $\beta$ in 
figure~\ref{jets:charm}b and is compared with predictions based on parton 
densities 
extracted from 
$F_2^D$ by H1 \cite{H1:F2D3}. 
The predictions are scaled down by a factor of 20, illustrating
the size of the factorisation breaking effects. At least for 
$\beta \lapprox 0.3$, the data and prediction are also rather different in 
shape. 


Another recent measurement from CDF is 
the fraction of visible
beauty production that is attributable to diffraction,
which yields the result \cite{cdf:b}
\begin{eqnarray*}
  \frac{\sigma_{\bar{b} b}^{\rm diff}}{\sigma_{\bar{b} b}^{\rm incl}}
= 0.62 \ \pm 0.19 \ {\rm (stat.)} \ \pm \ 0.16 \ {\rm (syst.)} \ ,
\end{eqnarray*} 
whereas the predictions on the basis of diffractive parton densities extracted
from $ep$ data are at the level of $10 \%$.

In a complementary analysis, Alvero et al \cite{ACTW} have extracted 
diffractive parton distributions from $F_2^D$ and photoproduction dijet data
from HERA and made predictions for various Tevatron measurements. 
Similarly large discrepancies are found when predicting
the rate of $W$ and 
dijet production as components of the system $X$ in the process
$\bar{p} p \rightarrow 
X \! \! \! \stackrel{\hspace*{0.1cm} _{_{ (-)}}}{p}$ \cite{cdf:jets,d0:jets}, 
with even larger 
differences for dijet production in
the double pomeron exchange process $p \bar{p} \rightarrow 
p X \bar{p}$.




Something beyond the simplest Regge and QCD factorisation assumptions
is clearly required to
describe simultaneously diffractive data from 
HERA and the Tevatron. 
The pertinent question now is
whether it is possible to build a phenomenological model of this breakdown
of factorisation. One possibility is that where 
beam remnants are present on both sides of a rapidity gap, rescattering takes
place, tending to destroy the gap. 
A very interesting place to study this possibility
is in photoproduction, where both factorisable (direct photon)
and non-factorisable (resolved photon) interactions may be expected to be
present. A first study can be found in \cite{h1:jets}.

\section{Total Cross Sections}


Total hadron-hadron cross sections are well described over a 
very wide energy range by two component Regge fits \cite{dl:stot}, 
corresponding (via the optical theorem) to the
exchange of the pomeron and a sub-leading ($\rho$, $\omega$, $f$, $a$)
trajectory in the elastic amplitude. The intercept of the leading pomeron
trajectory is most accurately determined from the high energy
rise in the $\bar{p} p$ cross section. Other total
cross sections such as $\pi^{\pm} p$ match this scheme well, though no data
exist at centre of mass energies $\gapprox 30 \ {\rm GeV}$. 

In the case where
one or both of the interacting hadrons is replaced by a photon, arguments
have been made that the presence of a bare photon coupling in addition to
the vector meson dominance hadronic component may lead to a faster rise of 
the total cross section with energy than is the case for pure hadron-hadron 
scattering. Eikonalised minijet models \cite{minijet}, incorporating 
semi-hard QCD interactions whilst avoiding the eventual violation of unitarity 
associated with simple Regge pole models, can be made to fit the available
data \cite{eikonal}. 
HERA data on the total $\gamma p$ cross section at centre of mass energy 
$W_{\gamma p} \sim 200 \ {\rm GeV}$ \cite{hera:stot} are consistent with the
simple Regge pole model, though the systematic errors are large and no
strong conclusion is yet possible. The
increasingly precise data from LEP on the total $\gamma \gamma$ cross section,
which may be expected to rise faster even than the $\gamma p$ cross section,
may shed some light on this issue.

\begin{figure}[htb] \unitlength 1mm
    \begin{center}
    \end{center}
    \begin{picture}(120,70)
       \put(0,-3){\epsfig{file=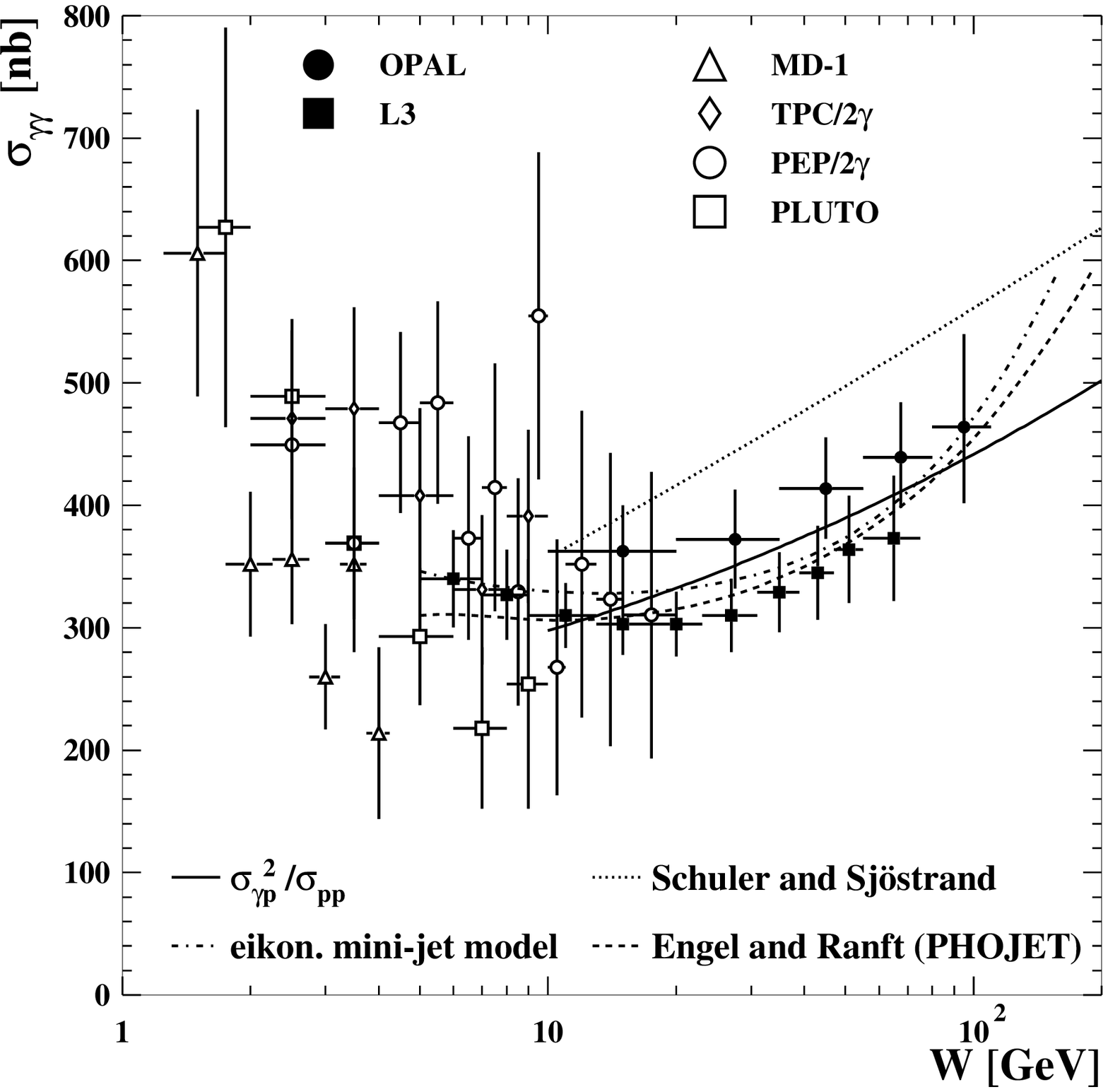,width=0.55\textwidth}}
       \put(80,-3){\epsfig{file=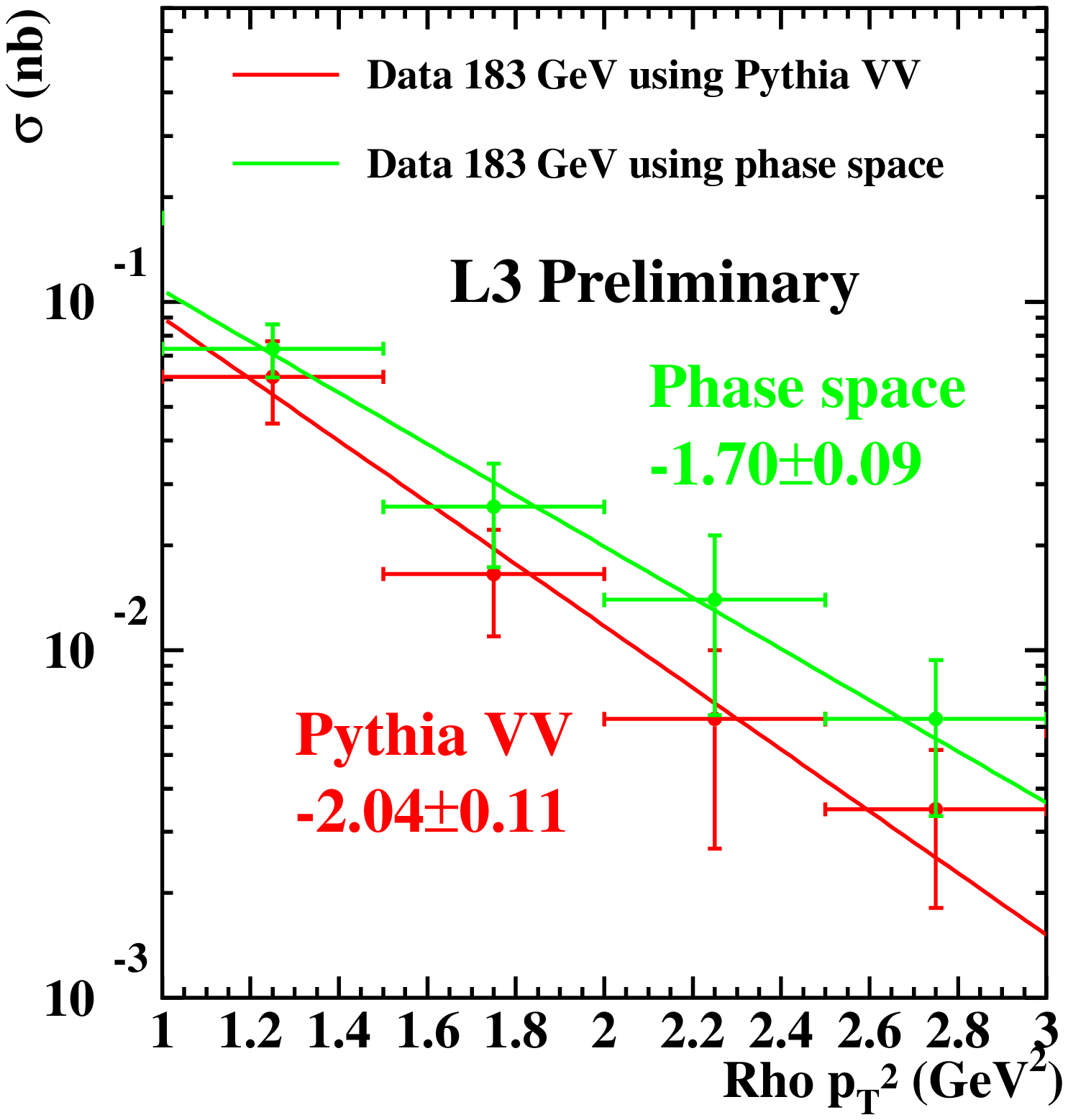,width=0.55\textwidth}}
       \put(65,22){\Large{\bf{(a)}}}
       \put(100,10){\Large{\bf{(b)}}}
    \end{picture}
    \caption{(a) A compilation of total $\gamma \gamma$ cross section
data. The L3 data are corrected using the PHOJET Monte Carlo model. The OPAL
data are the average of the values obtained when correcting with PHOJET or
PYTHIA, with a systematic error reflecting the difference between the two 
cases. (b) Distribution in $\pperp^2 (\rho) \simeq |t|$ 
for the `single dissociative' process 
$\gamma \gamma \rightarrow \rho^0 X$.}
    \label{gammagamma}
\end{figure}

Both L3 \cite{l3:stot} and OPAL \cite{opal:stot} have measured the total
$\gamma \gamma$ cross section in the region 
$10 \lapprox W_{\gamma \gamma} < 100 \ {\rm GeV}$. For the data used, both
electrons and many final state hadrons are lost down the
beam-pipe, making the kinematics
difficult to constrain. The data are shown, together with
lower energy fixed target data, in figure~\ref{gammagamma}a. The LEP data 
clearly show the high energy rise with $W$ observed in the
$\bar{p} p$, $pp$ and $\gamma p$ cross sections. A simple factorisation law
of the type $\sigma_{\gamma \gamma} = \sigma_{\gamma p}^2 / \sigma_{p p}$
describes the data remarkably well. 

It is not yet clear whether the rise
with $W$ is faster than that observed for total hadron-hadron
cross sections; the OPAL data are consistent with the pomeron intercept
describing soft hadronic interactions whereas the L3 result is significantly 
larger.
The results are
rather sensitive to the assumptions on $\alphareg(0)$. As can be seen
from figure~\ref{gammagamma}a, a model based on minijets \cite{eikonal} also
gives a reasonable description of the data, as do the 
Schuler and Sj\"{o}strand \cite{ss} and PHOJET \cite{phojet} models,
which attempt to make smooth transitions between the photon in
its hadronic and point-like manifestations. 

Improved data are
required before a firm conclusion can be reached concerning the possible 
anomalous behaviour of $\sigma_{\gamma \gamma}^{\rm tot}$. The main source
of error in the measurements arises from the model dependence of the acceptance
corrections, with results different at the level of $\sim 20 \%$ obtained
when PHOJET or PYTHIA \cite{pythia} 
is used for the corrections. The principal reason for
this is the different treatments of the diffractive
channels in the two models. Any constraints that can be placed on the 
diffractive processes in $\gamma \gamma$ scattering will improve the total
cross section measurement considerably.
Processes involving the
quasi-elastic production of vector mesons are likely to be the 
easiest to measure,
due to the well known decay angular distributions.
L3 have taken the first steps towards
measurements of the `quasi-elastic' ($\gamma \gamma \rightarrow \rho^0 \rho^0$)
and `single dissociation' ($\gamma \gamma \rightarrow \rho^0 X$) processes.
A measurement of the $t$ distribution of the single dissociation process is
shown in figure~\ref{gammagamma}b. Fitting the data to the usual
exponential parameterisation $d \sigma / d t \propto e^{b t}$ yields a 
slope parameter in the region $b \sim 2$. Information of this sort provides
very useful input to soft physics models and should ultimately 
reduce the model dependence uncertainties on the total cross section.

\section{Searches for BFKL Dynamics}

The BFKL 
evolution equation, which resums terms where 
large logarithms of the form $\ln 1/x$ multiply the coupling constant, 
must represent a valid approximation
to parton dynamics in some region of low $x$ phase 
space.
The search for 
evidence for BFKL behaviour is one of the principle current experimental 
activities in low $x$ physics. 
Although it has been shown that introducing
BFKL effects can improve the description of $F_2$ at low $x$ \cite{thorne},
it is not yet accepted that anything more than standard 
DGLAP 
evolution
is required to describe current inclusive DIS data. Exclusive final state 
measurements may ultimately produce the clearest BFKL signatures. Some of the
more promising areas of study are discussed below.


BFKL and DGLAP evolution have rather different implications for the details
of the parton ladder governing low $x$ DIS 
processes (figure ~\ref{diagrams}c). In the DGLAP case, one
expects an ordering in virtuality ($k_t$)
of the partons in the ladder, leading to
rapidity ordering of the transverse momenta of outgoing partons. The BFKL
scheme has no such strong ordering and therefore results in
anomalously large high $\ptrans$ 
hadron yields away from the photon vertex, for
example at central rapidity.
The central region of the $\gamma^* p$ frame corresponds to the forward
region of laboratory rapidity. 

\begin{figure}[htb] \unitlength 1mm
    \begin{center}
    \end{center}
    \begin{picture}(120,76)
       \put(-1,0){\epsfig{file=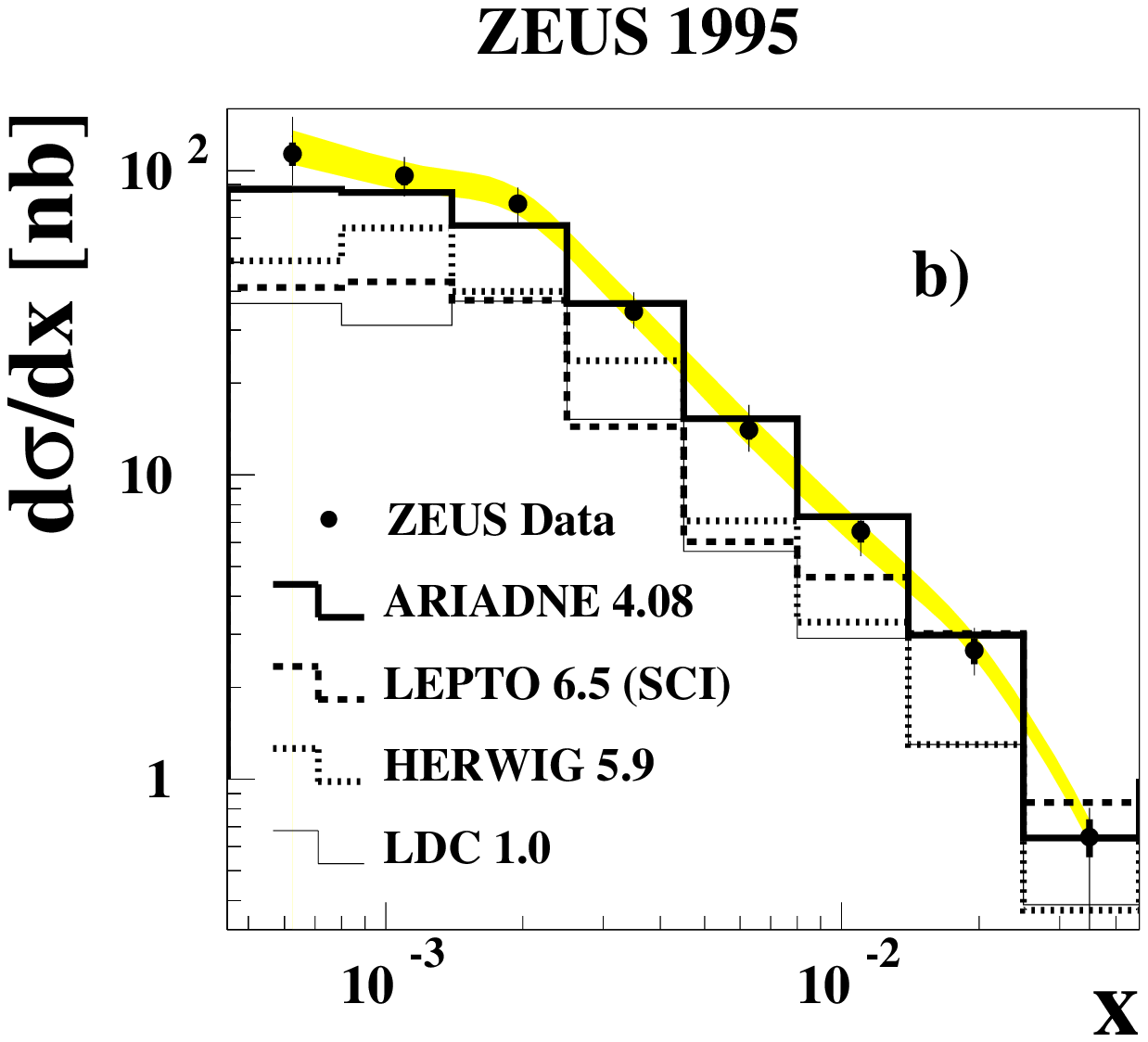,width=0.52\textwidth}}
       \put(76,0){\epsfig{file=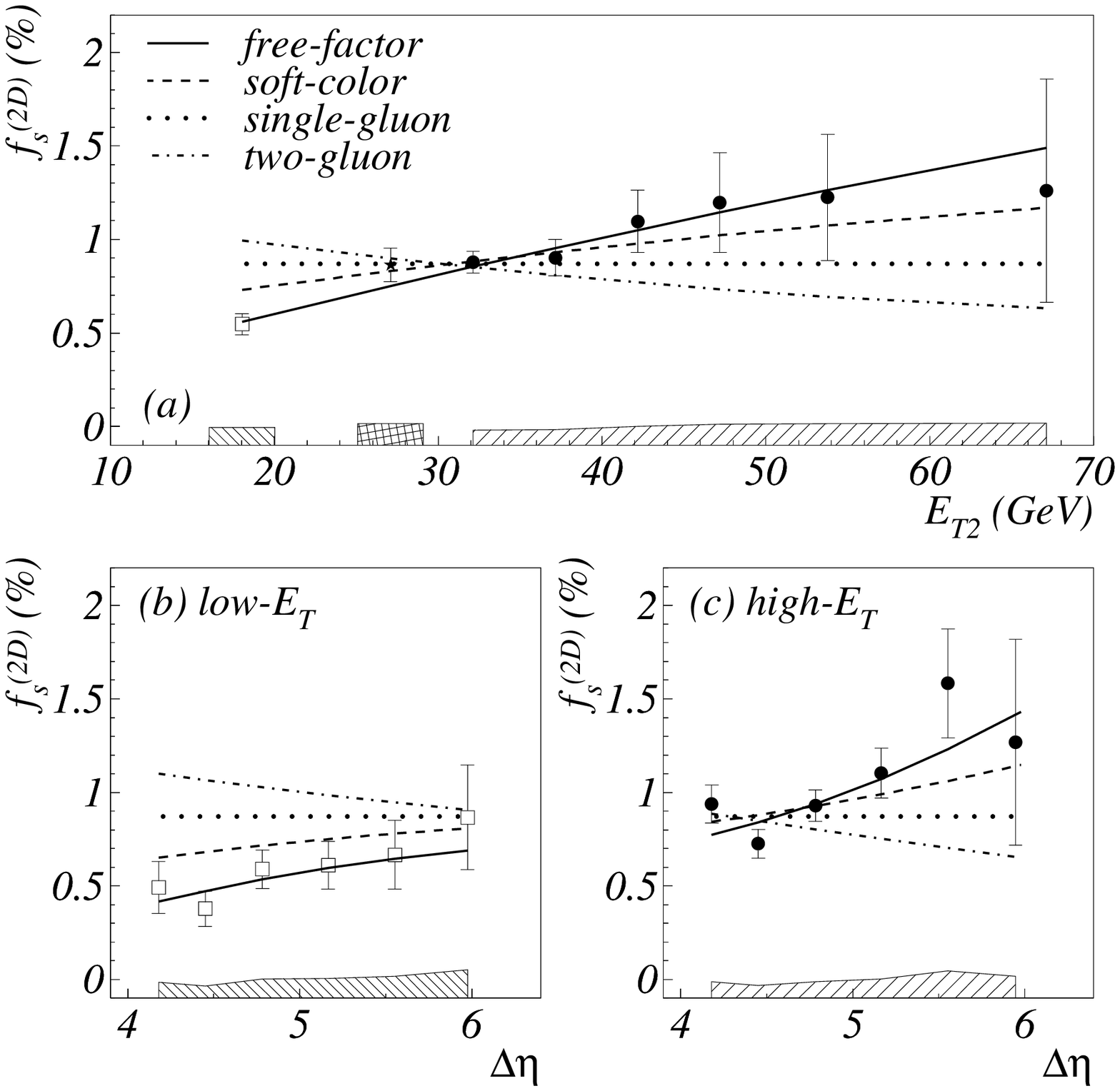,width=0.52\textwidth}}
       \put(56,61){\Large{\bf{(a)}}}
       \put(105,30){\Large{\bf{(b)}}}
    \end{picture}
    \caption{(a). Cross section for the production of jets with 
$E_{_T} > 5 \ {\rm GeV}$, $0.5 < E_{_T}^2 / Q^2 < 2$ and 
$p_z^{\rm jet} / E_p > 0.036$ in the Breit frame.
(b). Data on the fraction $f^{(2D)}_s$ of Tevatron dijet events where a
rapidity gap separates the jets at $\surd s = 1800 \ {\rm GeV}$.}
    \label{bfklfigs}
\end{figure}

Both H1 and ZEUS have studied the production of jets in this
difficult forward 
region \cite{fjets}. H1 have also measured the cross section for high 
$\ptrans$ forward $\pi^0$ production \cite{fpi}. 
Similar conclusions are reached in each case.
The ZEUS forward jet data
are shown in figure~\ref{bfklfigs}a. The data cannot be described by 
standard DGLAP models (labelled LEPTO and HERWIG). Only models that do not
impose strong transverse momentum ordering are able to describe the
data. One example is the ARIADNE model \cite{ariadne}, 
based on the colour dipole
model and simulating BFKL ordering. However, the lack of strong $k_{_T}$
ordering can also be modelled through the introduction of 
partonic structure to the virtual photon. This can be implemented 
in the RAPGAP \cite{rapgap}
Monte Carlo model, giving a successful description of all forward region data
produced at HERA to date.
Thus the final state data from the forward region at HERA
demonstrate that something more than the simplest DGLAP model of the low-$x$
parton ladder is required. However, resolved virtual photons provide an
alternative mechanism 
to BFKL to restore a good description of the data. Work on
events with large rapidity separations between pairs of jets, just beginning
at the Tevatron, may help to resolve some of these ambiguities. 

In appropriate kinematic regions, total, elastic and diffractive cross 
sections may all be describable in terms of the amplitude for elastic
parton-parton scattering via the exchange of gluon ladders, 
evolving according to BFKL dynamics. BFKL calculations are
most reliable where large scales are present at both vertices \cite{hit:bfkl}. 
One example is the total $\gamma^* \gamma^*$ cross section \cite{gammastar}.
Where both photons have sufficiently high virtuality, measurement conditions
at LEP are favourable and first data have appeared \cite{l3:gammastar}. 
The data
suggest a relatively strong energy dependence, which may be consistent
with BFKL predictions. However the present data can be described equally well 
by non-BFKL QCD models, for example those involving virtual photon
structure \cite{twogam}.

Another process where large scales are present at both vertices
is diffractive scattering at large $|t|$, where
precision data are starting to appear from HERA and the Tevatron. The 
quasi elastic process $\gamma p \rightarrow V Y$ where $V$ denotes a
vector meson and $Y$ is a proton or low mass proton excitation has been
measured for $V = J/\psi$, $\rho$ and $\phi$ \cite{hit:vm}. The results in the
relatively low $|t|$ regions accessed to date are mixed, only the $J/\psi$
fully conforming to the BFKL predictions.

The classic high $|t|$ diffractive process is the production of dijets 
separated by a rapidity gap, implying a net colour singlet exchange. 
Here, the magnitude of $t$ is close to the
jet $E_{_T}^2$ and is thus very large. The size of the cross section is 
usually quantified as the fraction of all
dijet events that have a rapidity gap between the jets.
Clear signals have been observed at large
jet pseudorapidity separation $\Delta \eta$ both in photoproduction
at HERA \cite{hera:jetgaps} and in $p \bar{p}$ interactions at the 
Tevatron \cite{tevatron:jetgaps,d0:jetgaps}. The gap fraction at large 
$\Delta \eta$ decreases with centre of mass energy, being
around 0.1 at $\surd s \simeq 200 \ {\rm GeV}$ at HERA, 0.025 at
$\surd s = 630 \ {\rm GeV}$ at the Tevatron and 0.01 at 
$\surd s = 1800 \ {\rm GeV}$ at the Tevatron. This trend is opposite to that
naively expected from BFKL calculations. However, it seems likely that
rapidity gap destruction due to reinteractions of beam remnants plays an
important role. 
Two very different models of these effects \cite{gapdie1,gapdie2} 
both predict
a rapidity gap survival probability that falls with centre of mass energy in a 
manner that qualitatively resembles that in the data.

The Tevatron gap fractions have been
measured as a function of jet $E_{_T}$ as well as 
$\Delta \eta$ (see figure~\ref{bfklfigs}b). The gap 
fraction is found to be flat or slowly rising with 
$E_{_T}^{\rm jet}$, which matches predictions based on the creation of 
rapidity gaps by soft colour interactions in otherwise standard dijet
events \cite{sci}. It has been
demonstrated that if rapidity gap destruction
effects are included, BFKL dynamics can also
describe these data \cite{gapdie2}.

All of the measurements discussed above can be interpreted 
in terms of  BFKL effects, yet none conclusively demonstrates the need for
BFKL at present colliders. Data from the upgraded Tevatron and HERA may
allow us to resolve this question.

\end{document}